\documentclass[aps,prapplied,preprint,superscriptaddress]{revtex4-2}

\usepackage{graphicx}
\usepackage[labelsep=endash, labelfont=sc]{caption}
\usepackage{subcaption}
\usepackage[export]{adjustbox}
\usepackage{hyperref}
\usepackage{float}
\usepackage[T1]{fontenc}
\usepackage[french, english]{babel}
\usepackage{csquotes}
\usepackage{siunitx}

\renewcommand{\selectlanguage}[1]{}

\begin{document}


\title{Helical coil design with controlled dispersion for bunching enhancement of the TNSA protons}


\author{A.~Hirsch-Passicos}
\email{arthur.emmanuel.hirsch@protonmail.com}
\affiliation{CEA-CESTA, Le Barp F-33114, France}
\affiliation{CELIA, University of Bordeaux-CNRS-CEA, UMR5107, Talence F-33405, France}
\author{C.L.C.~Lacoste}
\affiliation{CEA-CESTA, Le Barp F-33114, France}
\affiliation{CELIA, University of Bordeaux-CNRS-CEA, UMR5107, Talence F-33405, France}
\affiliation{INRS-EMT, Varennes (Québec), Canada}
\author{F.~André}
\affiliation{Thales AVS, 78 140 Vélizy-Villacoublay, France}
\author{Y.~Elskens}
\affiliation{Aix-Marseille université, PIIM, UMR 7345 CNRS, F-13397 Marseille, France}
\author{E.~D'Humières}
\affiliation{CELIA, University of Bordeaux-CNRS-CEA, UMR5107, Talence F-33405, France}
\author{V.~Tikhonchuk}
\affiliation{CELIA, University of Bordeaux-CNRS-CEA, UMR5107, Talence F-33405, France}
\affiliation{Extreme Light Infrastructure ERIC, ELI-Beamlines Facility, Dolní Brežany 25241, Czech Republic}
\author{M.~Bardon}
\affiliation{CEA-CESTA, Le Barp F-33114, France}
\affiliation{CELIA, University of Bordeaux-CNRS-CEA, UMR5107, Talence F-33405, France}



\date{\today}

\begin{abstract}
The quality of the proton beam produced by Target Normal Sheath Acceleration (TNSA) with high power lasers can be significantly improved with the use of helical coils. While they showed promising results in terms of focusing, their performances in terms of the of cut-off energy and bunching stay limited due to the dispersive nature of helical coils. A new scheme of helical coil with a tube surrounding the helix is introduced, and the first numerical simulations and an analytical model show a possibility of a drastic reduction of the current pulse dispersion for the parameters of high power laser facilities. The helical coils with tube strongly increase bunching, creating two collimated narrow-band proton beams from a broad and divergent TNSA distribution. The analytical model provides scaling of proton parameters as a function of laser facility features.
\end{abstract}


\maketitle

\section{Introduction}

Since the 2000s, ion acceleration by intense and short laser pulses \cite{Daido_2012, Macchi2012} 
has been a growing research field due to the very interesting properties of laser generated proton beams compared to ‘classically’ accelerated proton beams: short duration, high current, low emittance, high laminarity and high brightness \cite{Macchi2012, Roth2002, Borghesi2004, Cowan2004}. Several laser-driven ion acceleration processes are identified, such as the Target Normal Sheath Acceleration (TNSA) \cite{Snavely2000, Wilks2001}, radiation pressure acceleration \cite{sentoku2003high, schlegel2009relativistic} or collisionless shock acceleration \cite{martins2009ion}. TNSA is of particular interest, as it is the most robust scheme. It is characterized by a large angular divergence ($\sim$40°) \cite{Snavely2000} and an exponential energy distribution \cite{Snavely2000} with a cut-off energy depending on the laser intensity and energy \cite{Daido_2012, Macchi2012}. 

However, the large angular divergence and spectral distribution of TNSA ion beams are limitations for several potential applications, such as isochoric heating for warm dense matter (WDM) studies \cite{patel2003isochoric} and radio-isotope production for medical applications \cite{nemoto2001laser} or neutron production  \cite{lancaster2004characterization} for measurements of nuclear cross sections processes relevant to astrophysics. In order to improve the quality of TNSA ion beam, several schemes have been designed to focus, post-accelerate and select in energy ions such as the use of an active plasma lens \cite{yang2021designing}, magnetic self-focusing in a stack of conducting foils \cite{ni2013feasibility}, proximal target structures \cite{mcguffey2020focussing}, target curvature \cite{bin2009influence, carrie2011focusing, bartal2012focusing, patel2003isochoric, qiao2013dynamics} 
, target shaping \cite{ qiao2013dynamics, zakova2021improving}, solenoid field \cite{toncian2006ultrafast, harres2010beam} 
, ultra-thin foil targets \cite{kaymak2019boosted}, and nano and micro-structured targets \cite{giuffrida2017manipulation}. In particular, Kar et al. \cite{kar_guided_2016} developed a dynamic scheme for the focusing, post-acceleration and bunching of a TNSA proton beam. It consists in attaching a helical coil (HC) normally to the rear side of the target foil. The proton beam then propagates along the coil axis. This mechanism is the only one that allows to bunch, collimate and post-accelerate the proton beam at the same time, with a simple set-up as it uses only one laser beam.

This scheme modifies the TNSA proton beam distribution through several multi-physics and multi-scale processes: the target is charged positively during the escaping of the most energetic electrons during the TNSA process, this creates a discharge current through the helical coil attached to the rear side of the foil \cite{kar_guided_2016, Kar2016_2, AHMED2016172, ahmed__2017, aktan2019parametric}. The propagation of the discharge current through the helical coil induces an electromagnetic pulse (EMP) \cite{poye2015physics, consoli2020high} which can then focus, post-accelerate and energy select the TNSA protons. The helical coil geometry delays and guides both the current and electromagnetic pulse, matching them in speed and position with the proton beam, similarly to radiofrequency accelerators \cite{chick1951helix, chick1957experimental} and traveling wave tubes (TWT) \cite{pierce1950traveling}. TWT are used in the industry as amplificators: a continuous RF signal is sent at an extremity of the helix. A continuous beam of mono-energetic electrons generated from the cathode travels along the longitudinal axis at a speed slightly faster than the RF signal in the helix. The electrons are slowed down by the fields inside the helix and the transferred energy amplifies the RF signal at the exit of the helix. The HC scheme operates the inverse exchange where protons inside the HC gain energy from the fields generated by the current pulse.

This scheme has proven to be efficient experimentally in terms of focusing and spectral shaping of the TNSA proton beam \cite{kar_guided_2016, Kar2016_2, AHMED2016172, ahmed__2017, aktan2019parametric,bardon2020physics} but has also shown limitations in terms of the maximum energy of protons and of bunching. This is due to the dispersive nature of HC. Its impedance varies with the frequency, which modulates the current pulse intensity during its propagation along the helix. The current changes sign along the coil axis, inducing an alternance of accelerating and decelerating fields seen by the protons, leading to a reduction of the cut-off energy and bunching.

In order to go beyond this limitation, several schemes have been introduced:  Kar et al. use very short helical coils where the current does not have time to change sign \cite{kar_guided_2016}, Robertson et al. use two short helical coils in a row \cite{ferguson2023dual} and Liu et al. proposed a scheme with two sections of HC to skip phase reversal \cite{liu2022synchronous}.

In this article, we propose a new scheme of HC with tube, inspired by broadband high-power pulsed helix TWT \cite{freund1996theory, jiao2019key}. This scheme allows us to strongly improve the bunching effect of helical coils under the conditions relevant to 100 TW high intensity and short pulse laser facilities such as ALLS \cite{ALLS} and LULI2000 \cite{LULI2000}.

We first introduce this new scheme permitting to reduce drastically the dispersion inside the HC, as well as the physics of the dispersion of a current pulse inside a helical coil with tube. We then show the results of Particle-In-Cell (PIC) simulations and of a reduced model \cite{bardon2023DoPPLIGHT} of the current propagation in HC with and without tube under the conditions relevant to several laser facilities. Finally, we present the impact of a HC with tube on the TNSA spectrum and proton bunching.

\section{Description of a HC with tube}
\label{sec:Loaded_HC_Description}

\begin{figure}[H]
\centering
\begin{subfigure}{0.5\textwidth}
  \centering
  \includegraphics[width=1\textwidth]{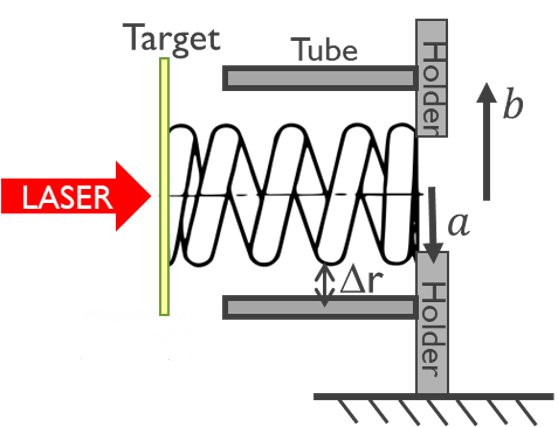}
  \caption{}
  \label{fig:Loaded_HC}
\end{subfigure}%
\begin{subfigure}{0.5\textwidth}
  \centering
  \includegraphics[width=1\textwidth]{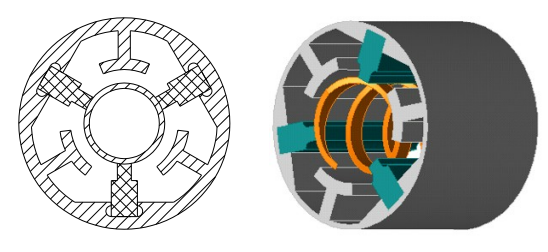}
  \caption{}
  \label{fig:Loaded_TWT}
\end{subfigure}
\caption{Scheme of : (A) helical coil with tube, (B) broadband high-power pulsed helix TWT \cite{jiao2019key}.}
\label{fig:Full_HC}
\end{figure}

The approach we propose to reduce the dispersion of helical coils consists in the use of helical coils with tube. In this scheme, shown in Figure \ref{fig:Loaded_HC}, the HC is inserted inside a metallic tube, the HC and the tube are both connected to the ground but the tube is not connected to the TNSA target, making the HC system similar to a coaxial line with a helical conductor inside. This is inspired by broadband high-power pulsed helix Traveling Wave Tubes (TWT), see Figure \ref{fig:Loaded_TWT}. The idea behind this scheme is to create a hybrid mode between the very dispersive helical coil and the dispersion-free conducting tube. The dielectric rod used in broadband high-power pulsed helix TWTs, has been suppressed due to the risk of short-circuit with the intense kA current going through the coil.

\subsection{Physics of the dispersion of a transient current in a HC with and without tube}
\label{sec:dispersion}

A model of current propagation was developed in the 1950s for travelling wave tubes (TWT) by Pierce \cite{pierce1950traveling} who proposed to approximate the helix wire by an infinitely thin cylinder with an anisotropic conductivity, which is non-zero only in the helical direction. The equations of this model were then developed into a circuit approach by Kino and Paik \cite{kino1962circuit}. This model is set in the Fourier domain, under the assumption that the pulse wavelength  along the longitudinal axis $z$ is larger than the turn length of the helix. It gives the following dispersion relation:

\begin{center}
\begin{equation}
    \omega= \frac{kc}{\sqrt{1+\cot^2(\Psi) \frac{I_1(\alpha)K_1(\alpha)}{I_0(\alpha)K_0(\alpha)}}}
    \label{eq:omega_HC}
\end{equation}    
\end{center}

\noindent where $I_{i}$ and $K_{i}$ are the modified Bessel functions of order $i$, $\alpha(\omega)=a\sqrt{k^2-\omega^2/c^2}$, $\Psi=\arctan(\frac{h}{2 \pi a})$ is the helix angle, $a$ is the radius of the helix and $h$ is the pitch. 

In particular, in the low frequency limit, $(a \omega /c) \cot\Psi  \ll 1$, one has $\alpha  \ll 1$, so $I_1(\alpha) \approx \frac{\alpha}{2}$, $I_0(\alpha) \approx 1$, $K_0(\alpha) \approx -ln(\alpha)$, $K_1(\alpha) \approx \frac{1}{\alpha}$, and $k \approx \omega/c$.

So, at low frequency, the mode phase velocity is close to the light velocity. 

On the other hand, for $(a\omega/c) \cot\Psi  \gg 1$, we have $\alpha \gg 1$, $I_0(\alpha),I_1(\alpha) \approx \sqrt{\frac{2}{\pi \alpha}}e^{\alpha}$, $K_0(\alpha), K_1(\alpha) \approx \sqrt{\frac{\pi}{2 \alpha}}e^{\alpha}$, and finally:

\begin{center}
\begin{equation}
    \omega \approx \frac{kc}{\sqrt{1+\cot^2(\Psi)}}
\end{equation}    
\end{center}

So, at high frequency, the phase velocity is reduced to $\frac{c}{\sqrt{1+cot^2(\Psi)}}$, which is smaller than c and depends only  on the helix pitch and radius.\\
Thanks to this asymptotic estimate, we could deduce that, in the helix, the low frequency mode propagates faster than the high frequency mode.\\
In an attempt to reduce the dispersion, in the 1990s, Freund and al. proposed a new theory for an already existing TWT scheme adding a loss-free conducting wall (tube) of radius $b$ to enclose the HC \cite{freund1992self}. The hybrid model is based on the same assumption of an infinitesimally thin helix, with the current propagating in the helical direction, in the Fourier domain. The authors obtain another dispersion relation:

\begin{center}
\begin{equation}
    \omega= kc / \left( \sqrt{1+\cot^2(\Psi) \frac{I_1(\alpha) I_0(\alpha b/a) [K_1(\alpha b/a)I_1(\alpha)-I_1(\alpha b/a)K_1(\alpha)]}{I_0(\alpha) I_1(\alpha b/a) [K_0(\alpha b/a)I_0(\alpha)-I_0(\alpha b)K_0(\alpha)]}} \right)
    \label{eq:omega_Tube}
\end{equation}    
\end{center}

In particular, one returns to the dispersion relation \ref{eq:omega_HC} in the limit where $b \rightarrow \infty$ then $I_1(\alpha  b/a) \rightarrow I_0(\alpha  b/a)$, $K_1(\alpha  b/a) \rightarrow 0$, $K_0(\alpha  b) \rightarrow 0$. Conversely, in the limit $b=a$ the helix current is short circuited by the axial current in the tube, so $\omega/k=c$. Indeed, if $b \rightarrow \infty$ we have only the HC and if $b \rightarrow a$ we get a perfectly conducting tube without dispersion, because that case corresponds to replacing the HC by the tube. 

\begin{figure}[H]
\centering
\begin{subfigure}{0.49\textwidth}
  \centering
  \includegraphics[width=1\textwidth]{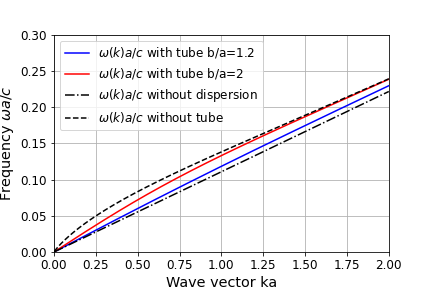}
  \caption{}
  \label{fig:KvsOmega}
\end{subfigure}
\begin{subfigure}{0.49\textwidth}
  \centering
  \includegraphics[width=1\textwidth]{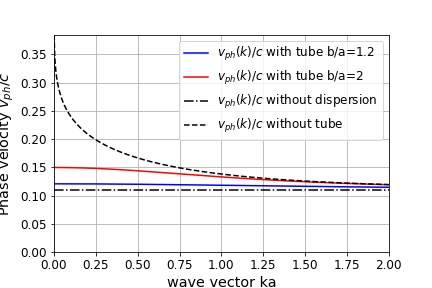}
  \caption{}
  \label{fig:kvsvph}
\end{subfigure}
\caption{Dependence of the current frequency (A) and phase velocity (B) on the wave number for a HC of radius $a=0.5$ mm, pitch $h=0.35$ mm, with tubes of different radii and without tube.}
\label{fig:Full_kvomega}
\end{figure}

We can conclude that the tube mitigates the current dispersion. This mitigation is presented in Figure \ref{fig:Full_kvomega} where the dispersion relation and the phase velocity $v_{ph}=\omega / k$ are plotted as a function of the wave vector for different values of $b/a$.

Therefore, the hybrid model combining a regular coil and a non-dispersive tube allows one to control the mode phase velocity and to reduce the dispersion. On the other hand, with the chosen tube geometry ($b/a=1.2$), the electric field available to accelerate the protons is reduced by a factor of about 1.5 compared to the free helix.

\section{Particle-In-Cell simulations and reduced model results for the current propagation of a TNSA shot in a HC with tube}

In order to validate the theoretical model, we first performed full-scale PIC simulations of TNSA protons with a HC surrounded by a metallic tube using SOPHIE \cite{cessenat2013sophie}, an electromagnetic particle-in-cell code developed at CEA-CESTA. It solves, in a self-consistent way, the Maxwell’s equations for the fields propagation in matter using boundary conditions (Perfect electric conductor, dielectric and magnetic materials) and the relativistic dynamics fundamental equation for the propagation of charged particles in vacuum. The geometry is modeled by appropriate boundary conditions at the perfectly conducting surfaces while the particles are injected following the given prescriptions of the temporal, angular and energetic distribution. The particle dynamics, target charging and discharge current propagation are self-consistently simulated.

\begin{figure}[H]
\centering
\includegraphics[width=1\textwidth]{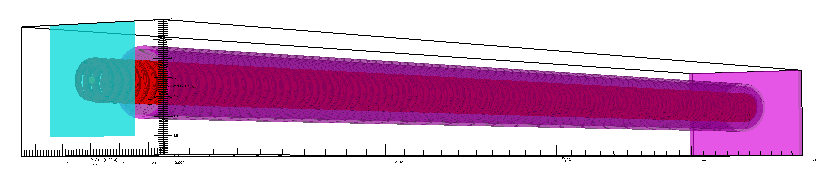}
\caption{Mesh of a HC with tube in SOPHIE. The target is in cyan, the helical coil in red and the target holder and tube in purple.}
\label{fig:Mesh}
\end{figure}

The solid structures in this simulation (target foil, coil, tube and grounds) are modelled with a resolution of $\Delta x=\Delta y=\Delta z=20$ $\mu$m as shown in Figure \ref{fig:Mesh}. In this section we present the results for a coil of length $L=40$ mm, a coil radius $a=0.5$ mm, a pitch $h=0.35$ mm and a wire diameter of $e=0.2$ mm. The tube is a perfectly conducting cylinder starting 2 mm after the target foil (in cyan In Figure \ref{fig:Mesh}), internal radius $b=0.9$ mm (i.e. $\Delta r=0.3$ mm between the HC and the conducting tube), and thickness $th=0.1$ mm. They are connected together by a perfectly conducting slab of 3.5 mm by 5 mm and of a thickness of $w=0.3$ mm at the end opposite to the TNSA target.

The simulations have been performed for two different laser parameters, one corresponding to the ALLS laser facility at INRS \cite{ALLS}, and the other one corresponding to the PICO2000 laser beam at LULI2000 \cite{LULI2000}. In each case, the emitted particles are separated into two different populations: the first population consists of protons, whose charge and energy distribution are taken from experimental data \cite{bardon2020physics, fourmaux2013investigation}, which also provide us the co-moving electron charge distribution. The second population consists of fast electrons, whose charge and temporal distributions are calculated from the numerical model ChoCoLaT \cite{poye2015dynamic}, developed by Poyé et al., with the laser parameters of each facility. We assumed the same temporal distribution for protons and electrons. The energy spectra of ejected particles are defined as follows:

\begin{table}[H]
    \centering
    \begin{tabular}{|c|c|c|}
    \hline
         & ALLS & PICO200 \\
    \hline
        Energy distribution & Maxwellian, $T=0.9$ MeV & Maxwellian, $T= 2.9$ MeV \\
    \hline
        Energy range & $1$ MeV$<E<6$ MeV & $1$ MeV$<E<19$ MeV \\
    \hline
        Temporal distribution & Gaussian, $\tau=3$ ps & Gaussian, $\tau=10$ ps \\
    \hline
        Proton charge & $Q_p=12$ nC & $Q_p=175$ nC \\
    \hline
        Fast electron charge & $Q_e=160$ nC & $Q_e=600$ nC \\
    \hline
    \end{tabular}
    \caption{Physical parameters of the ALLS and PICO2000 TNSA proton and fast electron populations.}
    \label{tab:my_label}
\end{table}

In both cases, the proton and co-moving electron angular distribution has the shape of a super-Gaussian function, defined by the function $\exp\left(-\frac{1}{2}\left(\theta / \theta_p\right)^{10}\right)$, centred around the longitudinal axis and defined by $\theta_p=19^{\circ}$ while the fast electron population is isotropically ejected in a solid angle of 2$\pi$.

\subsection{PIC Simulations}

\begin{figure}[H]
\centering
\begin{subfigure}{0.49\textwidth}
  \centering
  \includegraphics[width=1\textwidth]{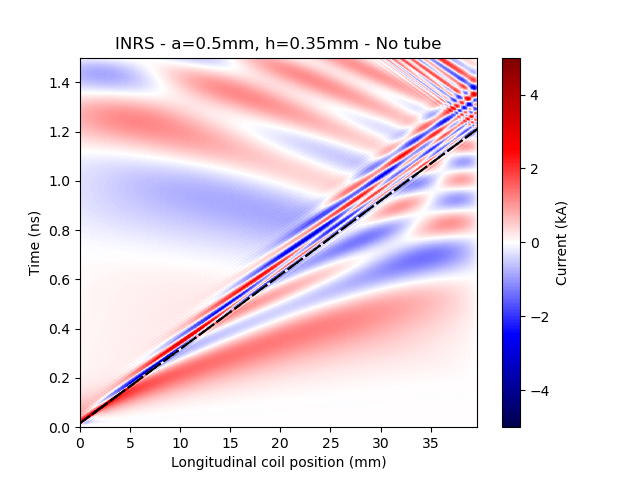}
  \caption{ }
  \label{fig:Coil_INRS_Current}
\end{subfigure}
\begin{subfigure}{0.49\textwidth}
  \centering
  \includegraphics[width=1\textwidth]{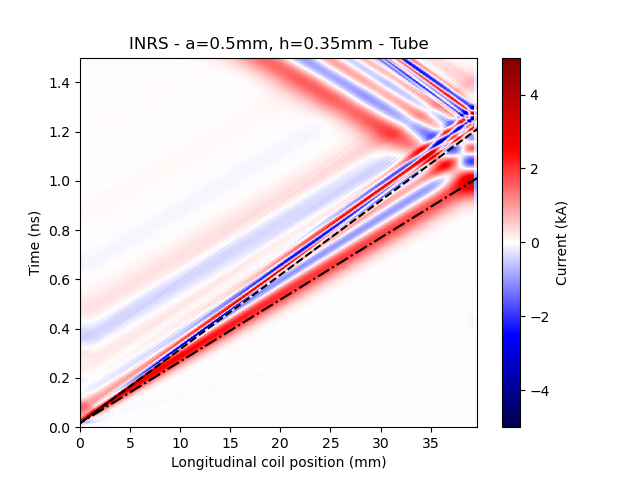}
  \caption{}
  \label{fig:Coil_Tube_INRS_Current}
\end{subfigure}
\caption{PIC simulation of the current pulse intensity in kA for an helical coil (A) without tube and (B) with tube as a function of time and along the HC axis. The particles features are defined in Table \ref{tab:my_label} for ALLS. The HC parameters are: length $L=40$ mm, radius $a=0.5$ mm, wire thickness of $0.02$ mm and pitch $h=0.35$ mm with a tube of radius $b=0.9$ mm. The dashed line corresponds to $V_{HC}$, the geometrical speed of the HC, the dash-dotted line corresponds to $V=1.2~V_{HC}$.}
\label{fig:Full_Current_INRS}
\end{figure}

The results of the SOPHIE simulations are presented in Figure \ref{fig:Full_Current_INRS}, which represents the current pulse propagation along the HC axis as a function of time. The impact of the metallic tube around the HC leading to a drastic reduction of the current dispersion can be seen from comparison of the left and right panels.

With a single HC, as seen in Figure \ref{fig:Coil_INRS_Current}, we observe many sign alternations of the current pulse propagating along the axis with the phase speed $V_{HC}$, corresponding to the longitudinal speed of a current pulse going at the speed $c$ along the helical coil: $V_{HC} = hc / \sqrt{h^2+4\pi^2a^2}$. One can see accelerated parts of the pulse propagating faster than $V_{HC}$, larger in time and lower in amplitude. All these features are clear signs of the coil's dispersive nature.

On the other hand, in the case of a HC with tube, one can clearly see a pulse of positive current propagating at a constant speed $V=1.2~V_{HC}$ and spreading temporally by a factor between 2 and 3 depending on the geometry, along with a reduction of amplitude by the same factor. This figure shows a clear reduction of the dispersion in the case of a HC with tube.

\begin{figure}[H]
\centering
\begin{subfigure}{0.49\textwidth}
  \centering
  \includegraphics[width=1\textwidth]{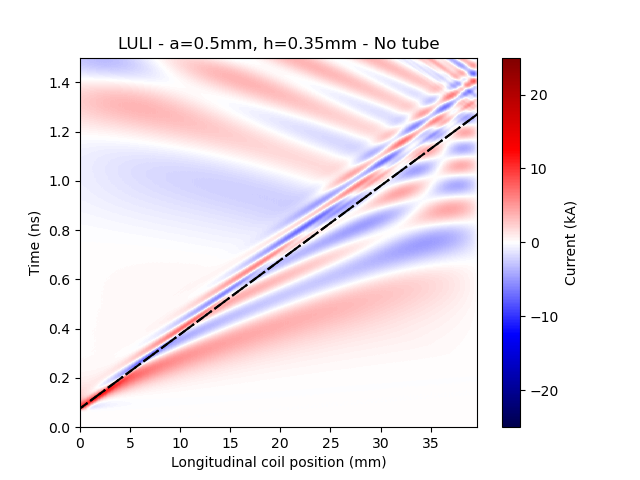}
  \caption{ }
  \label{fig:Coil_LULI_Current}
\end{subfigure}
\begin{subfigure}{0.49\textwidth}
  \centering
  \includegraphics[width=1\textwidth]{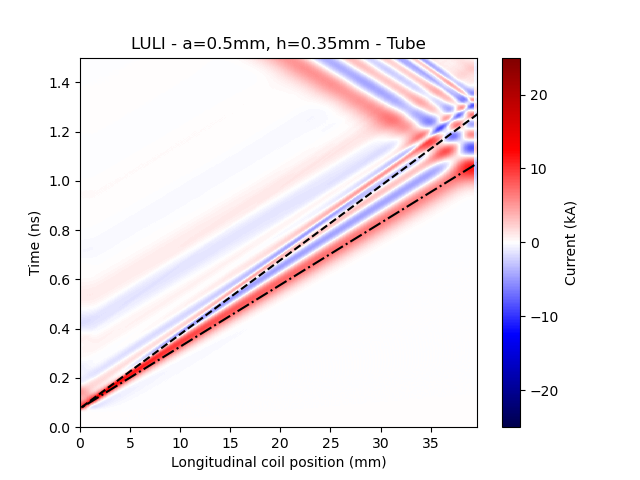}
  \caption{}
  \label{fig:Coil_Tube_LULI_Current}
\end{subfigure}
\caption{PIC simulation of the current pulse intensity in kA for an helical coil (A) without tube and (B) with tube as a function of time and along the HC axis. The particles features are defined in Table \ref{tab:my_label} for LULI2000. The HC parameters are: length $L=40$ mm, radius $a=0.5$ mm, wire thickness of $0.02$ mm and pitch $h=0.35$ mm with a tube of radius $b=0.9$ mm. The dashed line corresponds to $V_{HC}$, the geometrical speed of the HC, the dash-dotted line corresponds to $V=1.2~V_{HC}$.}
\label{fig:Full_Current_LULI}
\end{figure}

As one can see in Figure \ref{fig:Full_Current_LULI}, these results are very similar in the case of a higher energy facility, LULI2000. In the case of a HC with tube the positive current pulse propagates at a constant speed $V=1.2~V_{HC}$ and presents the same temporal spread and amplitude reduction as in the case of the ALLS laser facility.

\subsection{Reduced model results: the DoPPLIGHT code}

\begin{figure}[H]
\centering
\begin{subfigure}{\textwidth}
  \centering
  \includegraphics[width=1\textwidth]{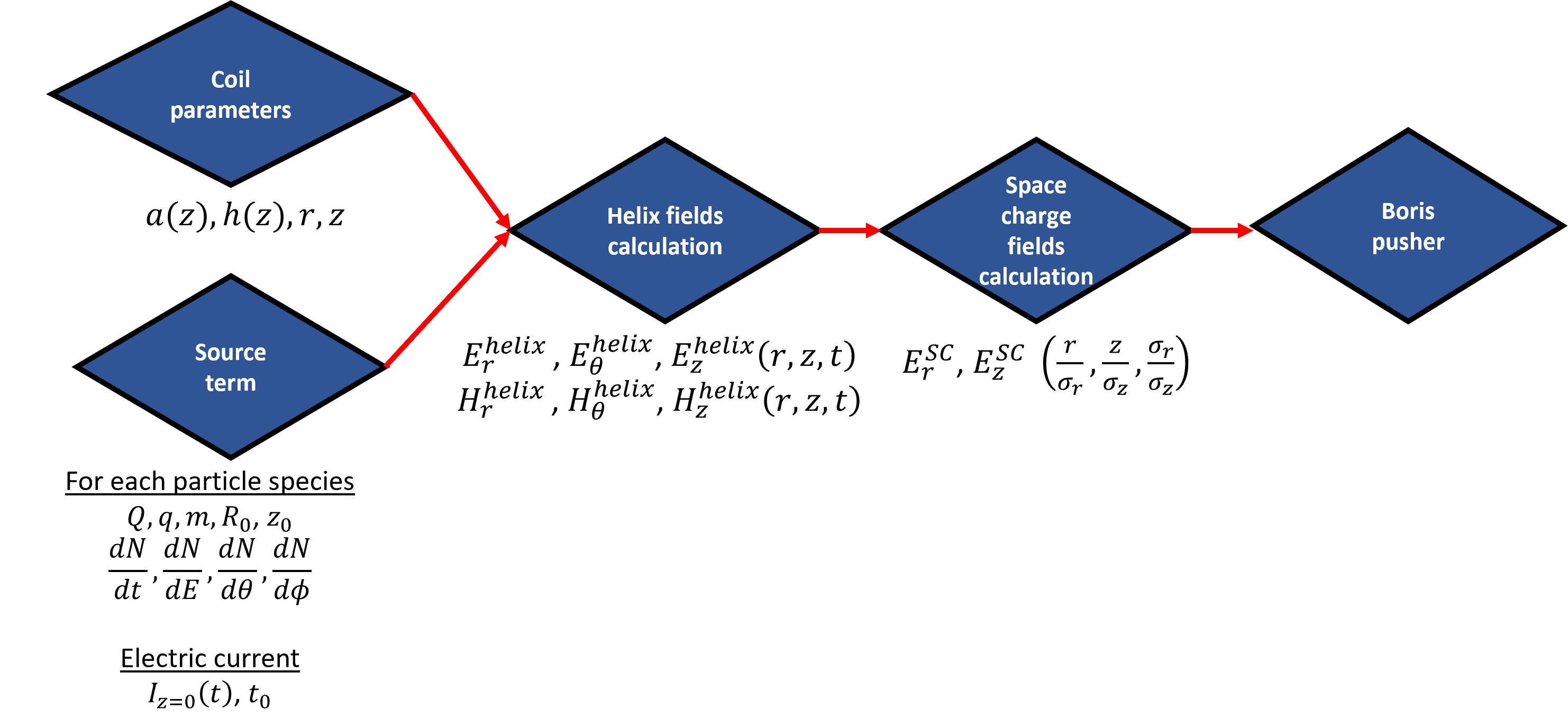}
  \caption{}
  \label{fig:DoPPLIGHT}
\end{subfigure}
\begin{subfigure}{\textwidth}
  \centering
  \includegraphics[width=1\textwidth]{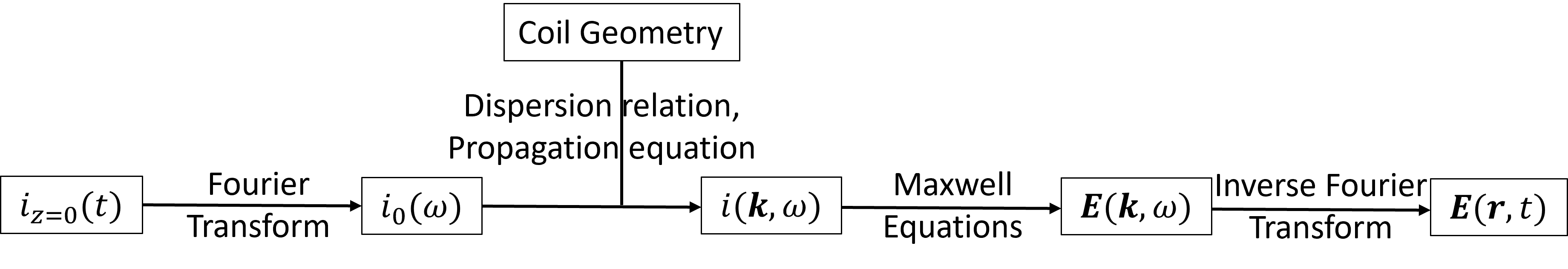}
  \caption{}
  \label{fig:Helix_Module}
\end{subfigure}
\caption{(A) Schematics of the model DoPPLIGHT (B) Schematics of the helix fields calculation module in DoPPLIGHT.}
\label{fig:Full_DoPP}
\end{figure}

The workflow of the model DoPPLIGHT \cite{bardon2023DoPPLIGHT} is presented in Figure \ref{fig:Full_DoPP}: it takes as input the helix geometry and the current at the beginning of the helix in the time domain, it then performs the Fourier transform of the current and uses it to calculate the fields according to Maxwell's equations. The boundary conditions are defined at the surface of the coil and tube. Following Pierce \cite{pierce1950traveling}, the coil is modeled as an infinitely thin cylinder with the current propagating in the helical direction. The tube is modeled as a perfect conductor. The fields are then transformed in the space-time domain by applying the inverse Fourier transform. The DoPPLIGHT model also calculates the space charge fields for a Gaussian shaped non-relativistic proton beam. The protons and electrons are then injected in the coil and their trajectories are calculated with the Boris pusher with the helix and space charge fields interpolated at each time step on the particle's position. Finally, at the exit of the coil, the code calculates the energy distribution of the protons. DoPPLIGHT operates in a 2D-axi-symmetric geometry; it is time-resolved and not a self-consistent model.

The PIC simulations are time consuming, but they can be used to validate the reduced numerical model of current propagation in the coil based on the dispersion equation described in Section \ref{sec:Loaded_HC_Description}. The code DOPPLIGHT \cite{bardon2023DoPPLIGHT} is developed specifically to describe the electric and magnetic fields produced in the helical coils. In the example presented below we consider a coil of length $L=40$ mm, radius $a=0.5$ mm and pitch $h=0.35$ mm, surrounded or not by a perfectly conducting tube of internal radius $b=0.8$ mm (i.e. $\Delta=0.3$ mm between the HC and the conducting tube).

The simulation box was discretised with $\Delta r=50$ $\mu$m and $\Delta z=100$ $\mu$m, with a time-step $\Delta t=0.4$ ps. The spectra of protons and electrons defined with the parameters from Table \ref{tab:my_label} and the current is defined analytically by a Gaussian function with full width at half maximum (FWHM) $\tau_{FWHM}=3$ ps and amplitude at $z=0$ mm $I_0=7$ kA for ALLS and by a Gaussian function with full width at half maximum (FWHM) $\tau_{FWHM}=8.5$ ps and amplitude at $z=0$ mm $I_0=30$ kA for LULI2000.

\begin{figure}[H]
\centering
\begin{subfigure}{0.49\textwidth}
  \centering
  \includegraphics[width=1\textwidth]{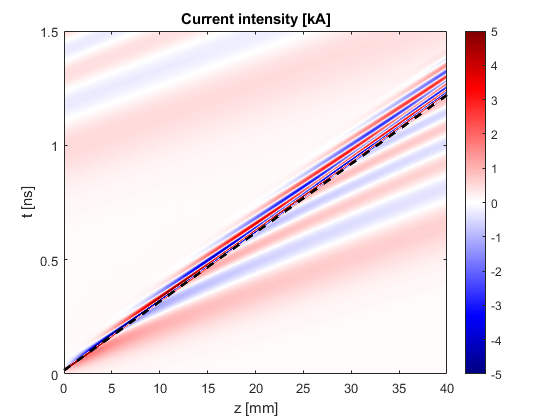}
  \caption{ }
  \label{fig:Coil_INRS_Current_DoPP}
\end{subfigure}
\begin{subfigure}{0.49\textwidth}
  \centering
  \includegraphics[width=1\textwidth]{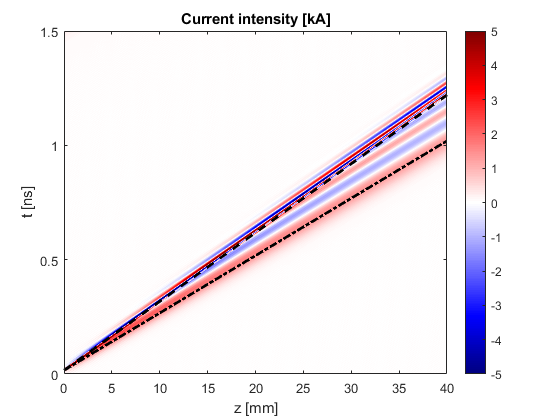}
  \caption{}
  \label{fig:Coil_Tube_INRS_Current_DoPP}
\end{subfigure}
\caption{DoPPLIGHT calculation of the current pulse intensity in kA for an helical coil (A) without tube and (B) with tube as a function of time and along the HC axis. The particles features are defined in Table \ref{tab:my_label} for ALLS. The current is defined analytically by a Gaussian function with FWHM $\tau_{FWHM}=3$ ps and amplitude at $z=0$ mm $I_0=7$ kA, delayed with respect to the particule emission by 6 ps. The HC parameters are: length $L=40$ mm, radius of the thin cylinder $a=0.5$ mm and pitch $h=0.35$ mm with a tube of radius $b=0.8$ mm. The dashed line corresponds to $V_{HC}$, the dash-dotted line corresponds to $V=1.2~V_{HC}$.}
\label{fig:Full_Current_INRS_DoPP}
\end{figure}

The results of the model are shown in Figures \ref{fig:Coil_INRS_Current_DoPP} and \ref{fig:Coil_Tube_INRS_Current_DoPP}. They are very similar to the PIC simulation results: we observe a positive pulse propagating at the same constant speed $V=1.2~V_{HC}$ with the same temporal spread by a factor of 2 and an associated amplitude reduction by the same factor than in the PIC simulations. 

These calculations are performed for several geometries and several laser source terms and we obtain in all cases good agreements between PIC simulations and DoPPLIGHT calculations as can be seen in the supplementary materials. We conclude that themodelisation of HC with tube in DoPPLIGHT is valid.

\begin{figure}[H]
\centering
\begin{subfigure}{0.49\textwidth}
  \centering
  \includegraphics[width=1\textwidth]{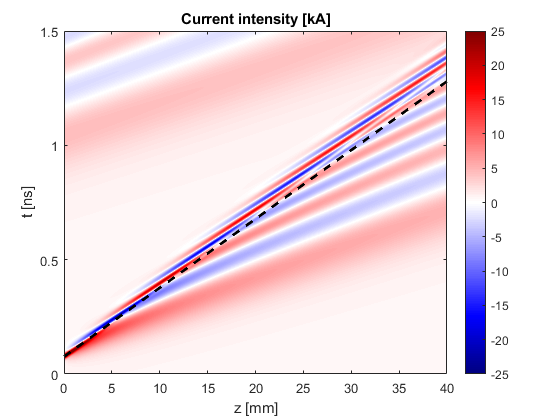}
  \caption{ }
  \label{fig:Coil_LULI_Current_DoPP}
\end{subfigure}
\begin{subfigure}{0.49\textwidth}
  \centering
  \includegraphics[width=1\textwidth]{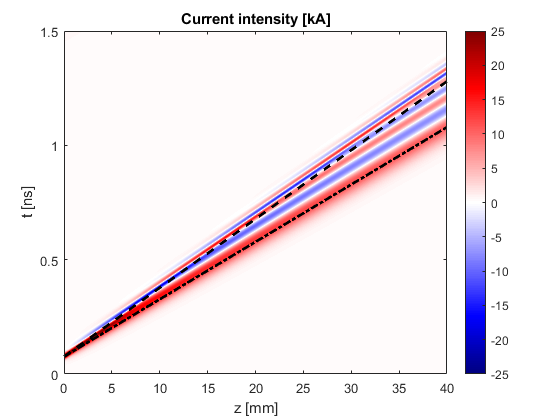}
  \caption{}
  \label{fig:Coil_Tube_LULI_Current_DoPP}
\end{subfigure}
\caption{DoPPLIGHT calculation of the current pulse intensity in kA for an helical coil (A) without tube and (B) with tube as a function of time and along the HC axis. The particles features are defined in Table \ref{tab:my_label} for LULI2000. The current is defined analytically by a Gaussian function with FWHM $\tau_{FWHM}=8.5$ ps and amplitude at $z=0$ mm $I_0=30$ kA, delayed with respect to the particule emission by 6 ps. The HC parameters are: length $L=40$ mm, radius of the thin cylinder $a=0.5$ mm and pitch $h=0.35$ mm with a tube of radius $b=0.8$ mm. The dashed line corresponds to $V_{HC}$, the dash-dotted line corresponds to $V=1.2~V_{HC}$.}
\label{fig:Full_Current_LULI_DoPP}
\end{figure}

As shown in Figure \ref{fig:Full_Current_LULI_DoPP}, these results are also valid when the source term is changed for the one of a higher energy facility, similar to LULI200.

\section{Effects of the HC with tube on the proton spectrum at the exit of the coil}

\subsection{Bunching effect}

\begin{figure}[H]
\centering
\includegraphics[width=0.8\textwidth]{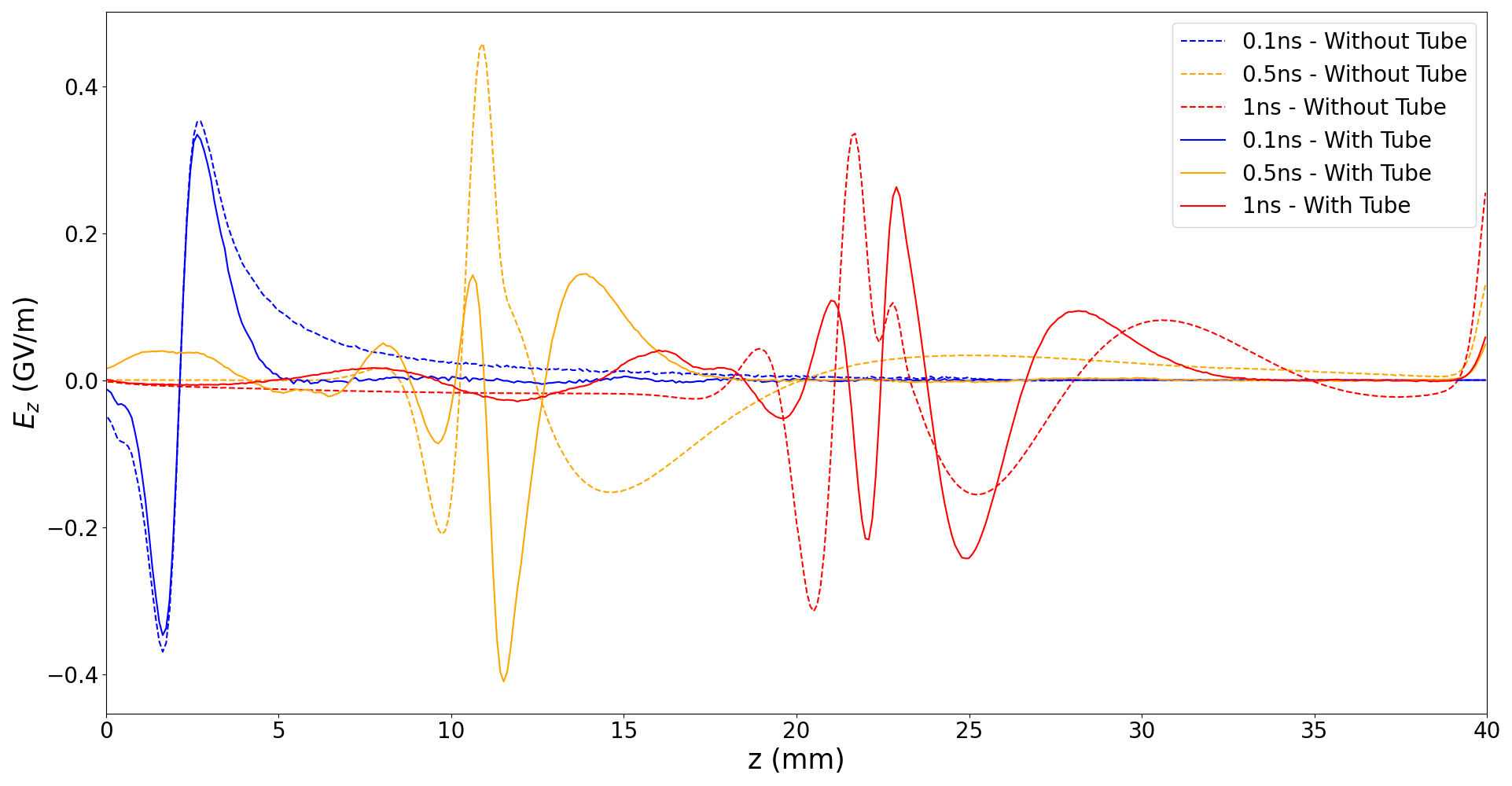}
\caption{Longitudinal field on the HC axis at different times, dashed lines correspond to an HC without tube, full lines to an HC with tube. The particles features set are defined in Table \ref{tab:my_label} for ALLS. The HC parameters are: length $L=40$ mm, radius $a=0.6$ mm, wire thickness of $0.02$ mm and pitch $h=0.35$ mm with a tube of radius $b=1$ mm.}
\label{fig:Ez}
\end{figure}

We studied the effects of electro-magnetic fields generated by the dispersion-reduced current pulse on the particle acceleration in the coil. Figure \ref{fig:Ez} presents the longitudinal field on the coil axis obtained in the simulations with the SOPHIE code at different times. We observe, for the HC with tube, a pulse with a positive longitudinal field followed by a pulse with a negative longitudinal field. The amplitude of these fields decreases with time (by a factor of 3 for the positive field), and their temporal spread increases (by a factor 1.5 for the positive field and by a factor of 2 for the negative field). A modulation appears in the tail of the pulse, but there are no sign changes and no acceleration of the pulse at the front like in the case of the HC without tube. This observation confirms that the dispersion is strongly reduced.

This structure of fields produces two important bunches of protons (see orange line in Figure \ref{fig:Spectre_Sophie}) around the characteristic energy of the HC $E_{HC} = 1/2 m_i V_{HC}^2$: one is more energetic, composed of the protons seeing the accelerating part of the field, another one is less energetic, composed of the protons under the influence of the decelerating field. It also produces a depletion zone situated in between the two bunches, where the proton population is several orders of magnitude less than the bunches. This feature is observed both in PIC simulations and in the reduced model DoPPLIGHT for several coil geometries with constant pitch and diameter, as can be seen in Figure \ref{fig:Full_Spectre} for one set of parameters:

\begin{figure}[H]
\centering
\begin{subfigure}{0.49\textwidth}
  \centering
  \includegraphics[width=1\textwidth]{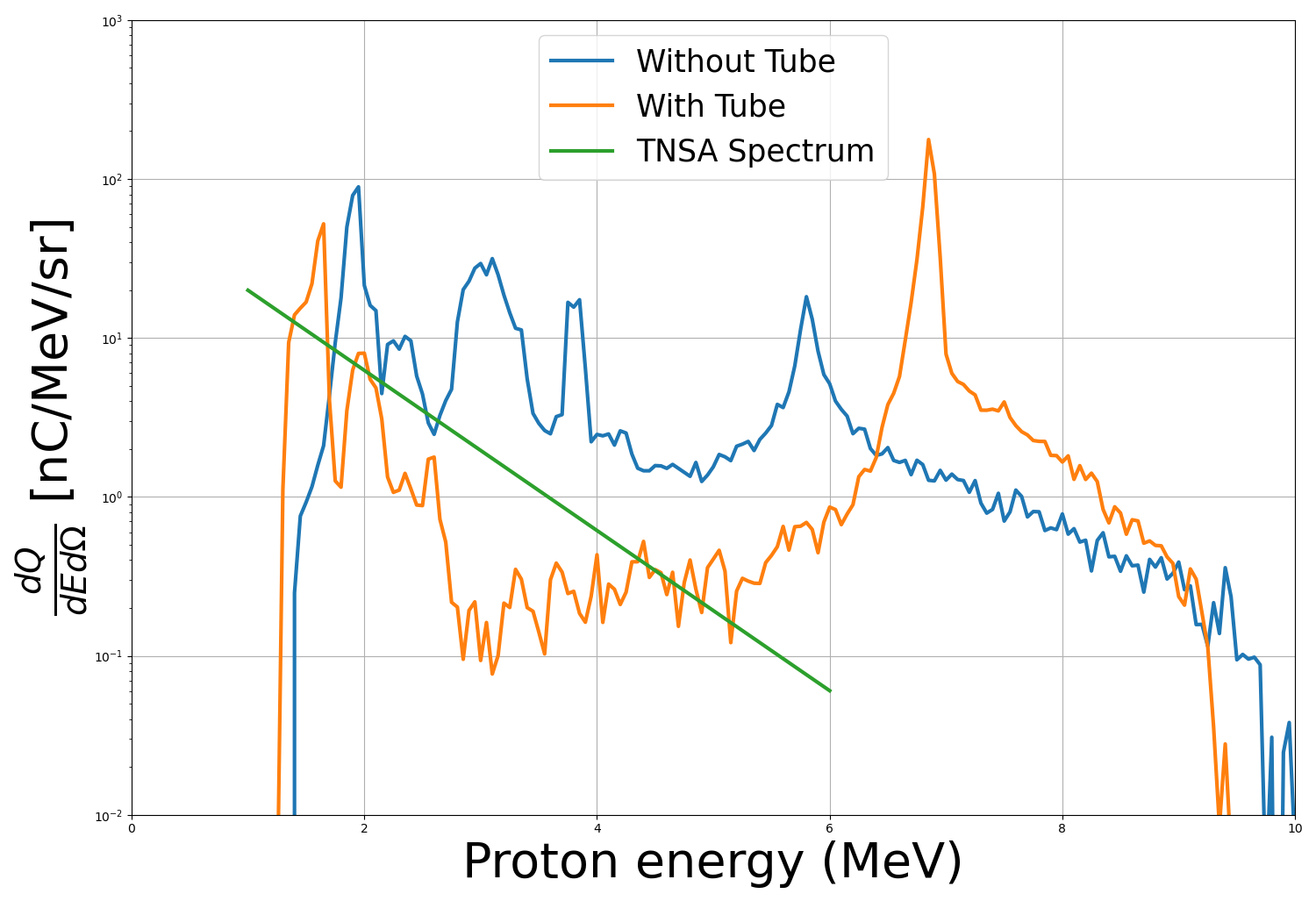}
  \caption{ }
  \label{fig:Spectre_Sophie}
\end{subfigure}
\begin{subfigure}{0.49\textwidth}
  \centering
  \includegraphics[width=1\textwidth]{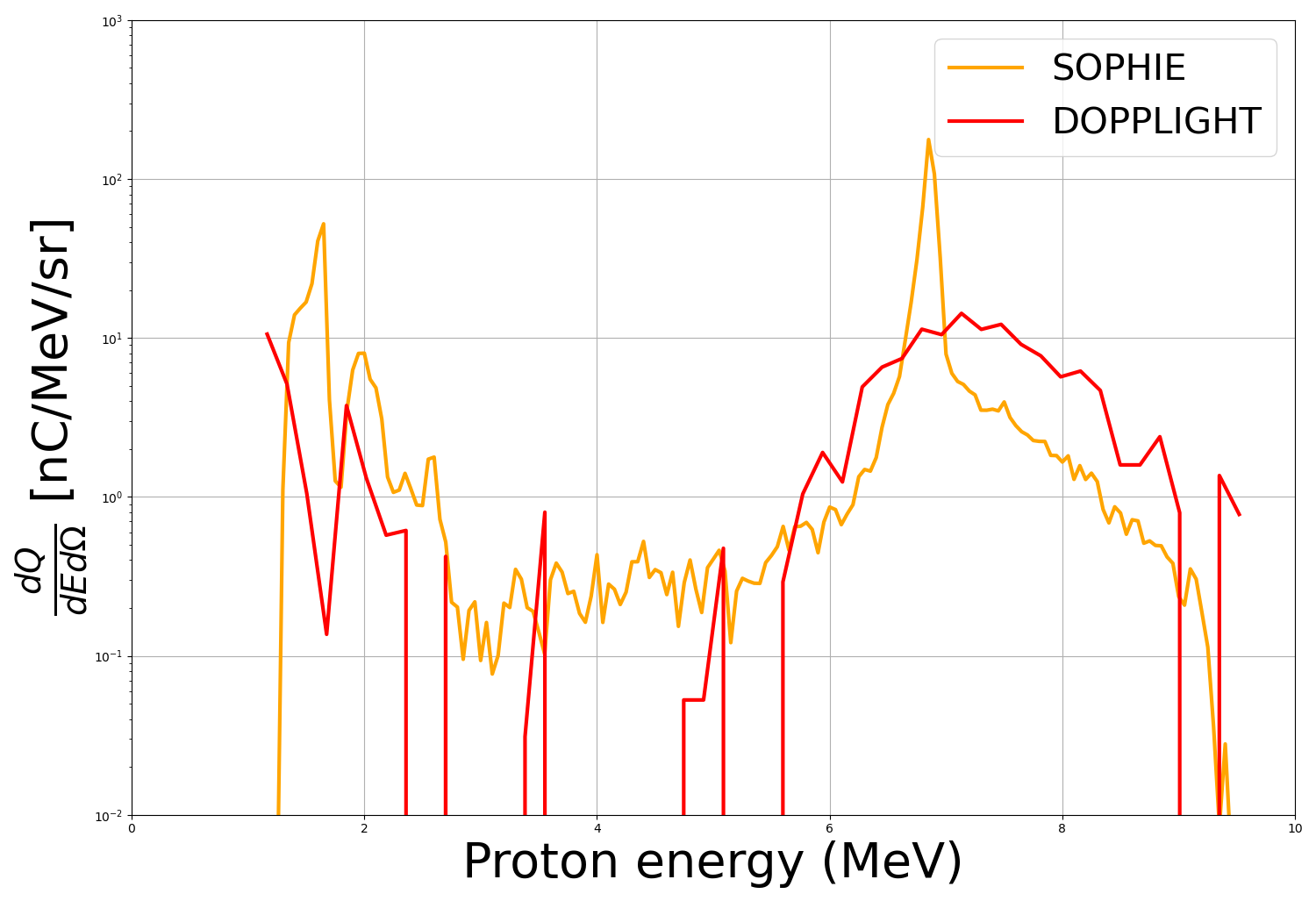}
  \caption{}
  \label{fig:DoppvsSoph}
\end{subfigure}
\caption{Spectra of accelerated protons: (A) TNSA spectrum (green line), spectrum at the exit of an HC without tube (blue line) and spectrum at the exit of an HC with tube (orange line) in PIC simulations, (B) spectrum at the exit of an HC with tube in PIC simulations (orange line) and in the numerical model DoPPLIGHT (red line). The particles features are in Table \ref{tab:my_label} for ALLS. The HC parameters are: length $L=40$ mm, radius $a=0.5$ mm, wire thickness of $0.02$ mm and pitch $h=0.35$ mm with a tube of radius $b=0.9$ mm, the HC characteristic energy is 2.9 MeV.}
\label{fig:Full_Spectre}
\end{figure}

Spectra obtained with tubes show a significant increase of the proton bunching due to the electromagnetic configuration produced by the tube's addition. The goal now is to be able to generate bunches at predicted energies.

We notice that this particular scheme of HC with tube keeps the focusing effect of the HC without tube, as we observe a stronger focusing of the proton bunch in the high energy population, by a factor between 10 and 100 depending on the geometry. The focusing is in agreement with experiments and with numerical simulations reported in previous works \cite{bardon2020physics}.

\subsection{Scaling of the bunching}

It would be interesting to design helical coils that produce bunches at predicted energies. We studied the correlation of bunch energy with the HC characteristic energy that depends only on the helix geometric parameters. For that, we made several calculations using DoPPLIGHT for different laser source terms, corresponding to ALLS and LULI2000, with radii going from $a=0.5$ mm to $a=0.8$ mm and pitches going from $h=0.3$ mm to $h=0.8$ mm, i.e. geometries that we can already manufacture. In all calculations, the radius of the tube was $b=a+0.3$ mm. We then normalised the characteristic energies of the bunches and of the depletion zone as well as the HC characteristic energy $E_HC$ by the input proton spectrum cut-off energy in order to get energy scalings independent of the laser facility.

\begin{figure}[H]
\centering
\begin{subfigure}{0.49\textwidth}
  \centering
  \includegraphics[width=1\textwidth]{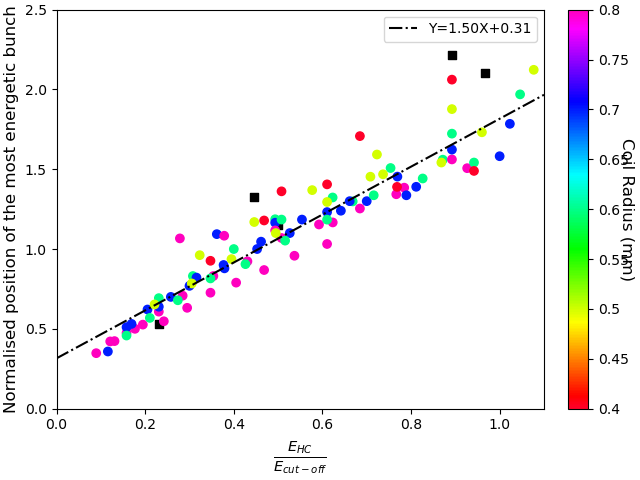}
  \caption{ }
  \label{fig:Scaling_max}
\end{subfigure}
\begin{subfigure}{0.49\textwidth}
  \centering
  \includegraphics[width=1\textwidth]{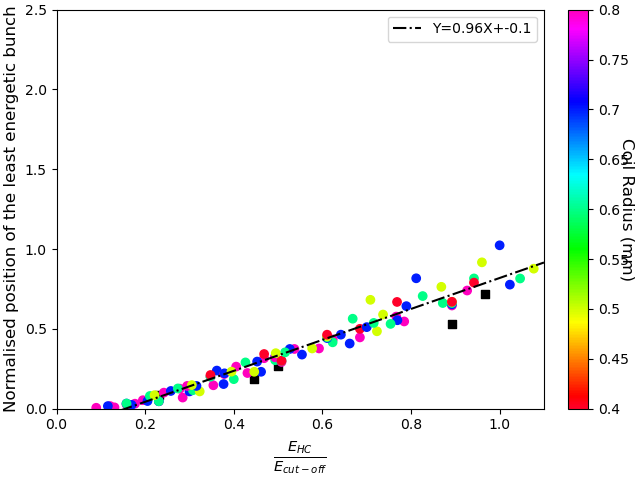}
  \caption{}
  \label{fig:Scaling_min}
\end{subfigure}
\begin{subfigure}{0.49\textwidth}
  \centering
  \includegraphics[width=1\textwidth]{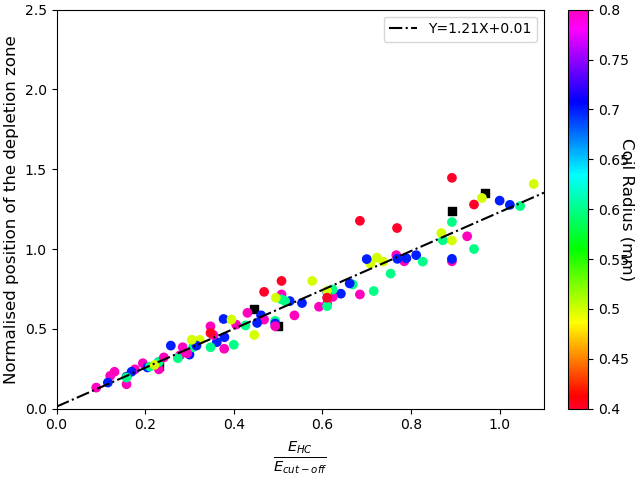}
  \caption{}
  \label{fig:Scaling_Creux}
\end{subfigure}
\caption{Scaling of the energy of the bunch with respect to the HC characteristic energy normalized by the cut-off energy: (A) normalised energy of the most energetic bunch, (B) normalised energy of the least energetic bunch, (C) normalised energy of the depletion zone. Colored dots correspond to DoPPLIGHT calculations, black dots to PIC simulations. Color code shows dependence on the coil radius. All HC are with tube.}
\label{fig:Scaling_Normal}
\end{figure}

We notice that all these scalings are independent of the coil radius and only depend on the normalised HC’s characteristic energy, making it the discriminatory parameter. 

As shown in Figure \ref{fig:Scaling_Normal} the normalised energy of the bunches and of the depletion zone depend linearly on the normalised HC characteristic energy.

The depletion zone energy corresponds to the velocity $V=1.2~V_{HC}$, i.e. the velocity of the positive current pulse observed in previous figures, the least energetic bunch energy corresponds to $V=1V_{HC}$ and the most energetic bunch corresponds to $V=1.5V_{HC}$. These respective deceleration and acceleration are consistent with the two longitudinal fields seen by the proton beam: the first one, shown in Figure \ref{fig:Ez}, is created by the constant positive pulse and the second one is created by the proton space charge field. Having similar shapes, they both accelerate the most energetic protons at the front and decelerate the less energetic protons at the rear.

\begin{figure}[H]
\centering
\includegraphics[width=0.6\textwidth]{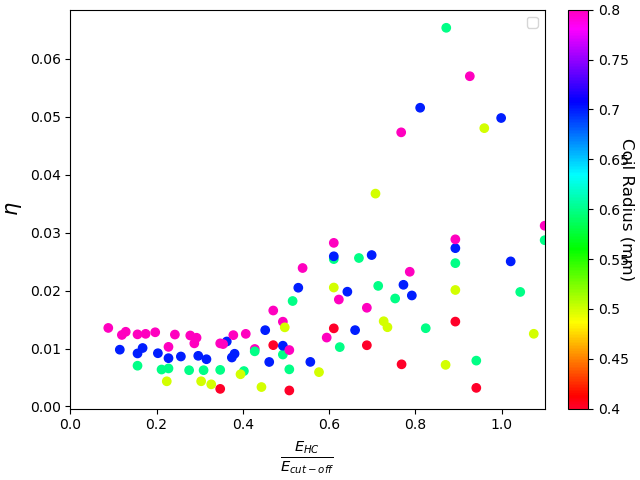}
\caption{Yield of protons at the exit of HC with tube as a function of the HC characteristic energy divided by the input cut-off energy; color code shows dependence on the coil radius.}
\label{fig:eta}
\end{figure}

We analyzed the yield of HC with tube, that is, the ratio of the charge at the exit of HC and the charge of the input proton distribution, as a function of HC characteristic energy normalized by the input cut-off energy and as a function of the coil radius. For $E_{HC}/E_{cutoff}<0.5$ the yield is constant and only depends on the coil radius, a larger radius corresponds to a higher yield. We then see an increase to an optimal yield value for $E_{HC}/E_{cutoff} \approx 0.8$ before decreasing to the previous level when the pulse becomes faster than the most energetic proton and does not impact the proton population anymore.

\begin{figure}[H]
\centering
\includegraphics[width=0.6\textwidth]{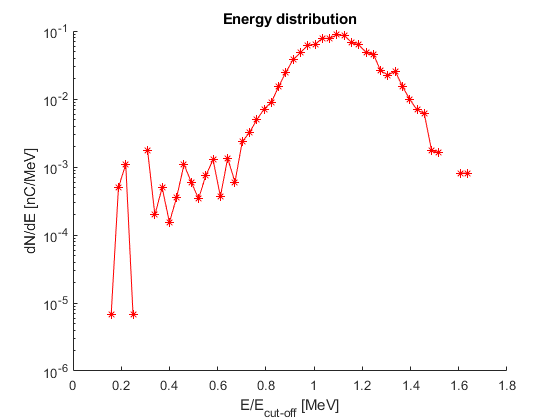}
\caption{Proton energy spectrum at the exit of a cone of radius $a=0.5$ mm and length $L=40$ mm calculated with DoPPLIGHT with only space charge fields for an ALLS-based input spectrum, as defined in Table \ref{tab:my_label}}
\label{fig:SC_Spectrum}
\end{figure}

This yield enhancement is due to the dominant space charge radial field at early times. Indeed the space charge is stronger where the proton density is higher and the unmodified TNSA proton distribution is composed mostly of low energy protons. As the HC fields are defocusing electrons from the beam and protons are spatially spread according to their energy, the radial space charge fields become dominant and proportional to the proton density. At the position of denser low energy protons, the HC field cannot compensate it even if its geometric speed is synchronised with these protons, and this leads to the loss of a large number of the low energy protons at the entrance of the HC. This can be understood from Figure \ref{fig:SC_Spectrum}, where the spectrum at the exit of an HC is shown with only the space charge fields acting on the input spectrum. We see that the space charge effects strongly defocus the low energy protons, leaving only the protons of energy $E>0.5E_{cut-off}$.

\section{Conclusions}

Previous experimental and numerical works have shown the limitations of post-acceleration and bunching of TNSA proton beams in regular HC targets. This is explained by the current dispersion in the coil, creating an alternance of accelerating and decelerating fields seen by the proton population.

We introduce a new scheme of HC targets that allows to drastically reduce the current dispersion in the coil. This scheme is simple and relatively easy to implement experimentally as it consists of surrounding the HC by a metallic cylinder.

The reduced model DoPPLIGHT is revised in order to take into account the metallic cylinder effects. The model results are in agreement with the large-scale PIC simulations made with the SOPHIE code.

The numerical study, on both SOPHIE and DoPPLIGHT, shows a strong effect in terms of bunching of protons above and below the characteristic energy of the helical coil, while keeping the focusing effect on the proton beam, which was observed in previous works on regular HC targets.
       
Furthermore, we obtain a scaling of the bunches energy with the HC characteristic energy, which is independent of the energy distribution of injected protons and can be used to design targets for specific energy bunches. Bunching is a feature interesting for applications such as isochoric proton heating, in order to heat a material at a specific depth, or radio-isotope production, which necessitates proton bunches at specific energies for the production of specific reactions.

This new acceleration scheme will be tested on the ALLS facility at INRS where we expect to observe the bunching and to verify both the scaling and the yield optimum demonstrated in this article.

The HC with tube targets can be modified by introducing progressive pitches in order to increase the cut-off energy of the protons by accelerating the longitudinal fields in synchronisation with the proton velocity.
       
\newpage

\appendix*

\section{Supplementary materials}
\label{sec:Supplementary}

In this appendix, we show the results of PIC simulations and DoPPLIGHT results for the current dispersion in HC with and without tube for different geometries and laser parameters.

\begin{figure}[H]
\centering
\begin{subfigure}{0.49\textwidth}
  \centering
  \includegraphics[width=1\textwidth]{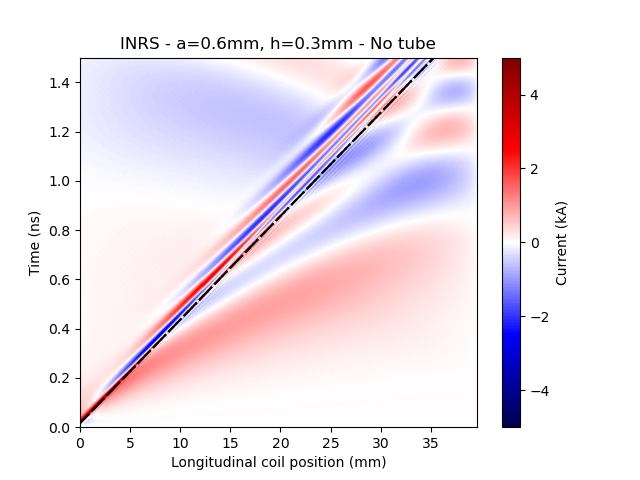}
  \caption{ }
  \label{fig:Coil_INRS_Current_A0P6_H0P3}
\end{subfigure}
\begin{subfigure}{0.49\textwidth}
  \centering
  \includegraphics[width=1\textwidth]{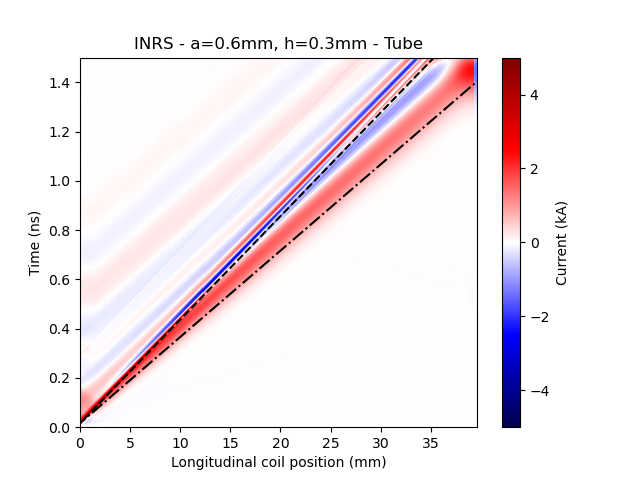}
  \caption{}
  \label{fig:Coil_Tube_INRS_Current_A0P6_H0P3}
\end{subfigure}
\caption{PIC simulation of the current pulse intensity in kA for an helical coil (A) without tube and (B) with tube as a function of time and along the HC axis. The particles features are defined in Table \ref{tab:my_label} for ALLS. The HC parameters are: length $L=40$ mm, radius in the heart of the coil $a=0.6$ mm, external radius=$0.7$ mm and step $h=0.3$ mm with a tube of radius $b=1$ mm. The dashed line corresponds to $V_{HC}$, the geometrical speed of the HC, the dash-dotted line corresponds to $V=1.2~V_{HC}$.}
\label{fig:Full_Current_INRS_A0P6_H0P3}
\end{figure}

\begin{figure}[H]
\centering
\begin{subfigure}{0.49\textwidth}
  \centering
  \includegraphics[width=1\textwidth]{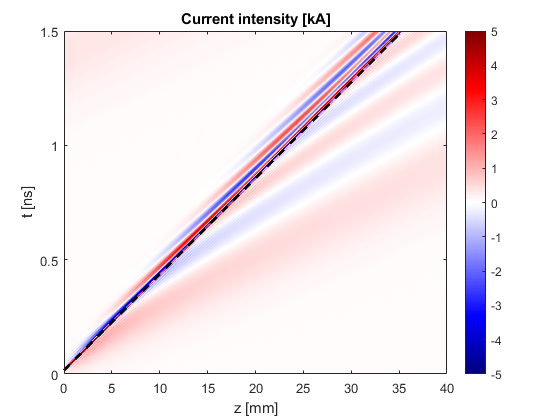}
  \caption{ }
  \label{fig:Coil_INRS_Current_DoPP_A0P6_H0P3}
\end{subfigure}
\begin{subfigure}{0.49\textwidth}
  \centering
  \includegraphics[width=1\textwidth]{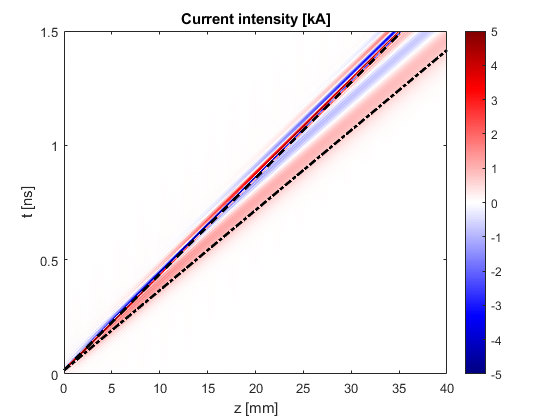}
  \caption{}
  \label{fig:Coil_Tube_INRS_Current_DoPP_A0P6_H0P3}
\end{subfigure}
\caption{DoPPLIGHT calculation of the current pulse intensity in kA for an helical coil (A) without tube and (B) with tube as a function of time and along the HC axis. The particles features are defined in Table \ref{tab:my_label} for ALLS. The current is defined analytically by a gaussian with FWHM $\tau_{FWHM}=3$ ps and amplitude at $z=0$ mm $I_0=7$ kA, delayed with respect to the particule emission by 6 ps. The HC parameters are: length $L=40$ mm, radius of the thin cylinder $a=0.6$ mm and step $h=0.3$ mm with a tube of radius $b=0.9$ mm. The dashed line corresponds to $V_{HC}$, the dash-dotted line corresponds to $V=1.2~V_{HC}$.}
\label{fig:Full_Current_INRS_DoPP_A0P6_H0P3}
\end{figure}

\begin{figure}[H]
\centering
\begin{subfigure}{0.49\textwidth}
  \centering
  \includegraphics[width=1\textwidth]{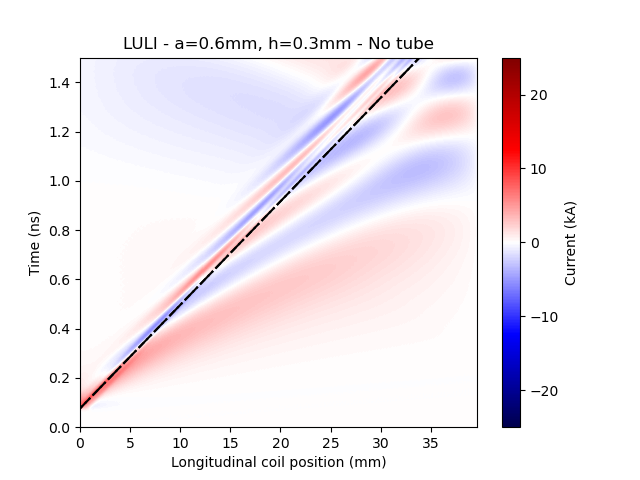}
  \caption{ }
  \label{fig:Coil_LULI_Current_A0P6_H0P3}
\end{subfigure}
\begin{subfigure}{0.49\textwidth}
  \centering
  \includegraphics[width=1\textwidth]{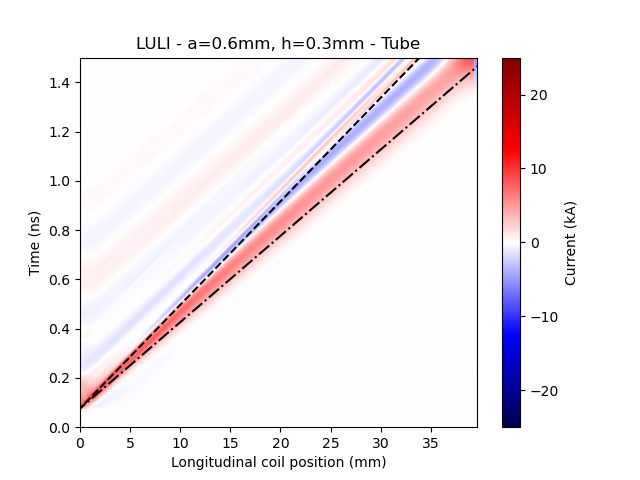}
  \caption{}
  \label{fig:Coil_Tube_LULI_Current_A0P6_H0P3}
\end{subfigure}
\caption{PIC simulation of the current pulse intensity in kA for an helical coil (A) without tube and (B) with tube as a function of time and along the HC axis. The particles features are defined in Table \ref{tab:my_label} for LULI. The HC parameters are: length $L=40$ mm, radius in the heart of the coil $a=0.6$ mm, external radius=$0.7$ mm and step $h=0.3$ mm with a tube of radius $b=1$ mm. The dashed line corresponds to $V_{HC}$, the geometrical speed of the HC, the dash-dotted line corresponds to $V=1.2~V_{HC}$.}
\label{fig:Full_Current_LULI_A0P6_H0P3}
\end{figure}

\begin{figure}[H]
\centering
\begin{subfigure}{0.49\textwidth}
  \centering
  \includegraphics[width=1\textwidth]{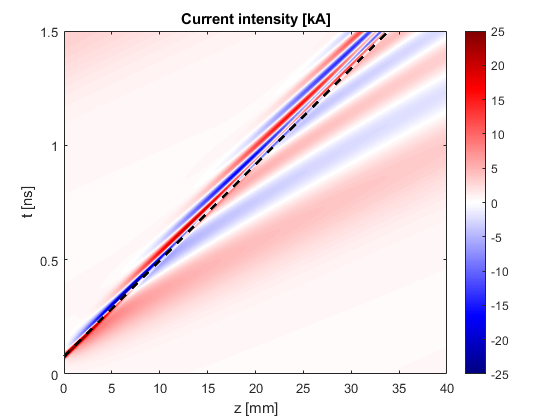}
  \caption{ }
  \label{fig:Coil_LULI_Current_DoPPA0P6_H0P3}
\end{subfigure}
\begin{subfigure}{0.49\textwidth}
  \centering
  \includegraphics[width=1\textwidth]{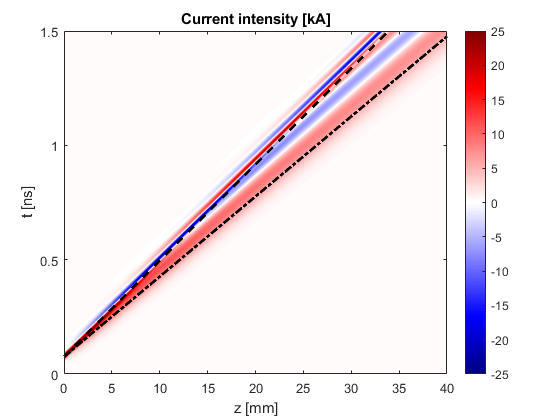}
  \caption{}
  \label{fig:Coil_Tube_LULI_Current_DoPPA0P6_H0P3}
\end{subfigure}
\caption{DoPPLIGHT calculation of the current pulse intensity in kA for an helical coil (A) without tube and (B) with tube as a function of time and along the HC axis. The particles features are defined in Table \ref{tab:my_label} for LULI. The current is defined analytically by a gaussian with FWHM $\tau_{FWHM}=8.5$ ps and amplitude at $z=0$ mm $I_0=30$ kA, delayed with respect to the particule emission by 6 ps. The HC parameters are: length $L=40$ mm, radius of the thin cylinder $a=0.6$ mm and step $h=0.3$ mm with a tube of radius $b=0.9$ mm. The dashed line corresponds to $V_{HC}$, the dash-dotted line corresponds to $V=1.2~V_{HC}$.}
\label{fig:Full_Current_LULI_DoPP_A0P6_H0P3}
\end{figure}

\begin{figure}[H]
\centering
\begin{subfigure}{0.49\textwidth}
  \centering
  \includegraphics[width=1\textwidth]{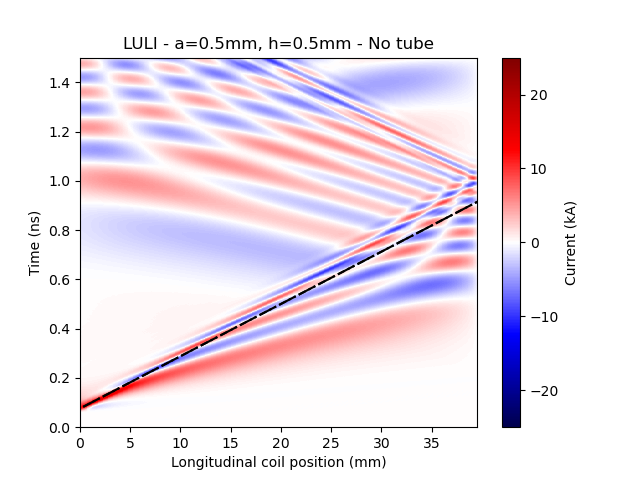}
  \caption{ }
  \label{fig:Coil_LULI_Current_A0P5_H0P5}
\end{subfigure}
\begin{subfigure}{0.49\textwidth}
  \centering
  \includegraphics[width=1\textwidth]{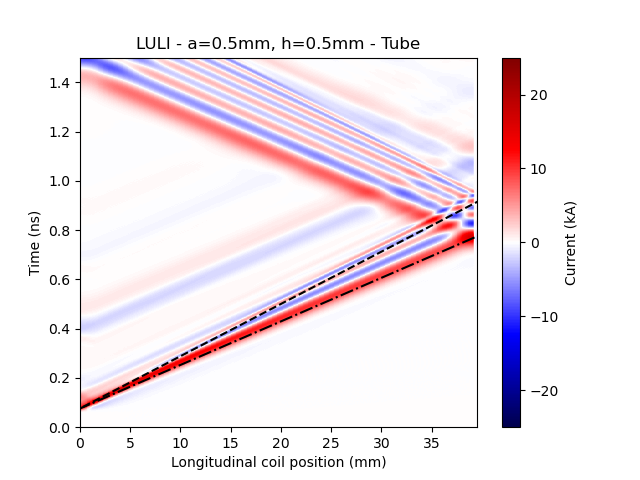}
  \caption{}
  \label{fig:Coil_Tube_LULI_Current_A0P5_H0P5}
\end{subfigure}
\caption{PIC simulation of the current pulse intensity in kA for an helical coil (A) without tube and (B) with tube as a function of time and along the HC axis. The particles features are defined in Table \ref{tab:my_label} for LULI. The HC parameters are: length $L=40$ mm, radius in the heart of the coil $a=0.5$ mm, external radius=$0.6$ mm and step $h=0.5$ mm with a tube of radius $b=0.9$ mm. The dashed line corresponds to $V_{HC}$, the geometrical speed of the HC, the dash-dotted line corresponds to $V=1.2~V_{HC}$.}
\label{fig:Full_Current_LULI_A0P5_H0P5}
\end{figure}

\begin{figure}[H]
\centering
\begin{subfigure}{0.49\textwidth}
  \centering
  \includegraphics[width=1\textwidth]{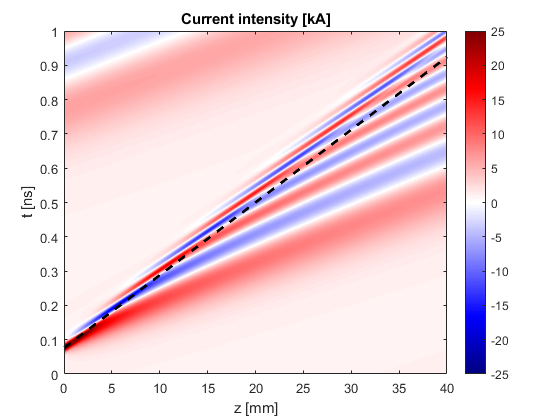}
  \caption{ }
  \label{fig:Coil_LULI_Current_DoPP_A0P5_H0P5}
\end{subfigure}
\begin{subfigure}{0.49\textwidth}
  \centering
  \includegraphics[width=1\textwidth]{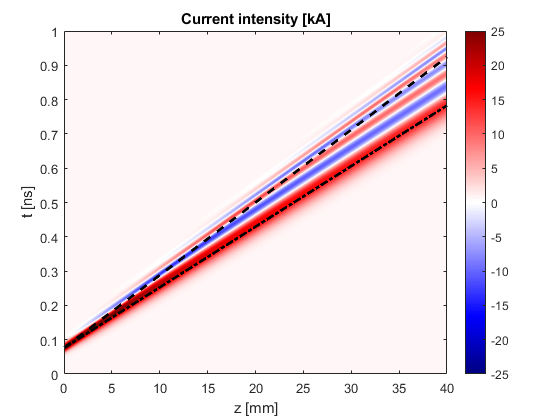}
  \caption{}
  \label{fig:Coil_Tube_LULI_Current_DoPP_A0P5_H0P5}
\end{subfigure}
\caption{DoPPLIGHT calculation of the current pulse intensity in kA for an helical coil (A) without tube and (B) with tube as a function of time and along the HC axis. The particles features are defined in Table \ref{tab:my_label} for LULI. The current is defined analytically by a gaussian with FWHM $\tau_{FWHM}=8.5$ ps and amplitude at $z=0$ mm $I_0=30$ kA, delayed with respect to the particule emission by 6 ps. The HC parameters are: length $L=40$ mm, radius of the thin cylinder $a=0.5$ mm and step $h=0.5$ mm with a tube of radius $b=0.8$ mm. The dashed line corresponds to $V_{HC}$, the dash-dotted line corresponds to $V=1.2~V_{HC}$.}
\label{fig:Full_Current_INRS_DoPP_A0P5_H0P5}
\end{figure}

\begin{acknowledgments}
The authors are grateful to K. Aliane , O. Cessenat, F. Doveil, D. Minenna and A. Poyé for their help and useful discussions.

This work is supported by the CEA/DAM laser-plasma experiments validation project and the CEA/DAM basic technical and scientific studies project. This work was granted access to the HPC resources of IRENE under the allocation No. A0130512899 made by GENCI
\end{acknowledgments}

\bibliography{biblio.bib}

\end{document}